\documentstyle[12pt]{article}
\begin{document}
\thispagestyle{empty}
\pagestyle{myheadings}\markright{arch-ive/9412002}
\renewcommand{\thefootnote}{\fnsymbol{footnote}}
\centerline{\bf MULTIPLE SCATTERING ANALYSIS OF}
\centerline{\bf CU--{\it K} EXAFS IN
$\rm Bi_2Sr_{1.5}Ca_{1.5}Cu_2O_{8+\delta}$\footnote[1]{to be published in:
Proc. Int. Workshop on ''Anharmonic Properties of High--$T_c$ Cuprates'',
Bled, Slovenia, Sep. 1-6, 1994 (World Scientific).}}
\vskip 1.1cm
\centerline{J. R\"ohler\footnote[2]{e-mail: abb12@rs1.rrz.uni-koeln.de}}
\centerline{\small\it Universit\"at zu K\"oln, Physikalisches Institut}
\centerline{\small\it D-50937 K\"oln, Germany}
\vskip 0.3cm
\centerline{and}
\vskip 0.3cm
\normalsize
\centerline{R. Cr\"usemann}
\centerline{\small\it Universit\"at Stuttgart, Institut f\"ur
Physikalische Chemie}
\centerline{\small\it D-70569 Stuttgart, Germany}
\vskip 0.8cm
\centerline{ABSTRACT}
\vskip 0.6cm
\leftskip 3pc
\rightskip 3pc
\footnotesize
\baselineskip 12pt
\noindent
We have analyzed the Cu {\it K}--EXAFS of $\rm
Bi_2Sr_{1.5}Ca_{1.5}Cu_2O_{8+\delta}$ using a full multiple scattering
analysis in a cluster with diameter $d\simeq 7.6$ {\AA}. The layered
structure has numerous quasi one-dimensional structural elements which give
rise to significant multiple scattering contributions in the EXAFS. We
 confirm the Sr/Ca ratio of the sample is 1:1, and one Ca atom is located
close to a nominal Sr-site. At 40 K the dimpling angle in the $\rm
CuO_2$\/-plane is found to be $\leq 3.5^{\circ}$.

\leftskip 0pc
\rightskip 0pc
\vskip 0.6cm
\normalsize
\leftline{\bf 1. Introduction}
\vskip 0.4cm
\noindent
Knowledge of the positions and the dynamics of the atoms forming the layer
between the doping block and the $\rm CuO_2$\/-planes in the high-$T_c$
cuprates is of particular interest for the understanding of their electronic
structure and mechanism of superconductivity \cite{Mue93}. The layer
contains divalent alkaline-earth atoms ($\rm Ba^{2+},\- Sr^{2+},\- Ca^{2+}$)
and the apical oxygen. The apical oxygen exhibits many anomalous features:
high anisotropic polarizability \cite{Mue90}, anharmonic motion, zero
contribution to the oxygen isotope effect \cite{ZecMue}. Linking the doped
$\rm CuO_2$ planes to the charge reservoir ({\it N.B.} one-dimensional or
quasi-one-dimensional structural Cu-, Bi-, Tl-, or HgO elements
\cite{EscDre}), the apical oxygen site is sensitive to the electronic
coupling of the 'active' layers, and unusual anharmonicities are expected.
The anharmonic potential of the apical oxygen was recently proposed to be of
a double well type \cite{MusTru}, but so far there is no unambigous
experimental evidence for that special type of anharmonicity
\cite{SteMae,Roe93}.

Less attention is attributed to the positions and the dynamics of the
alkaline-earth atoms located in the layer between chains and planes.
Usually these atoms are considered to act as mere spacers, which adjust
their positions dependent on the charge redistribution between the
electronically active layers. But there is increasing evidence for
electron-phonon coupling along the $c$\/-axis involving the whole
polarizable earth-alkaline--oxygen layer \cite{SchKal}.

This contribution deals with the determination of the local structure of the
polarizable layer in $\rm Bi_{2+x}(Sr,Ca)_3Cu_2O_{8+\delta}$ (the 2212
phase) by Cu-{\it K}\/ x-ray extended-absorption fine-structure (EXAFS). The
photoelectron interference patterns were analyzed using a full multiple
scattering calculation up to a total pathlength of 8 {\AA}. Advantageously
the 2212 phase exhibits a single Cu site. Therefore interference effects
between two local Cu substructures as {\it e.g} in $\rm
YBa_2Cu_3O_{7-\delta}$ are excluded. But the Bi-cuprates are structurally
more complex than the 123 system. For a review see {\it e.g.} Majewski
\cite{Maj} and references therein.

The 2212 phase exhibits an extended single-phase region with variable Ca,
Sr, Bi and oxygen content, which allows considerable variations from the
nominal stoichiometric composition $\rm Bi_2Sr_2Ca_1Cu_2O_8$. Systematic
investigations of the homogeneity range of the 2212 phase confirm the
nominal stoichiometric composition is not included in the single-phase
region; stoichiometric 2212 samples are always found to contain various
secondary Ca--Sr--Bi cuprates. Clearly, contributions from such possible
secondary phases superimpose the local structure of the primary phase, and
usually there is no reliable procedure to disentangle a complex structural
multiphase mixture.

Single-phase materials are therefore a necessary prerequisite for
investigations of local, ({\it i.e.} non translationally invariant)
deviations from the average crystallographic structure. Even single-phase
optimum doped high-$T_c$ materials, due to their inherent non-stoichiometry,
exhibit many vacancies, interstitials or atomic site changes. We have shown
recently by EXAFS that oxygen vacancies at the chain ends in single-phase
$\rm YBa_2Cu_3O_{7-\delta}$ ($\delta > 0.1$) create two-site apex
configurations superimposing possible dynamic double-well anharmonicities
\cite{Roe93}.

Local structural refinements lacking the constraints due to lattice symmetry
appear to be rather complex. Nevertheless, local lattice distortions arising
{}from the inherent nonstoichiometry of the high-$T_c$ cuprates are safely
detectable, as well as variations as a function of temperature, provided
well characterized single-phase materials have been selected.
\vskip 0.6cm
\leftline{\bf 2. Experiment}
\vskip 0.4cm
\leftline{\it 2.1 Sample characterization}
\vskip 0.4cm
\noindent
$T_c$ of the Bi 2212 phase varies between $\simeq 50$ K and 94 K and is a
function of the oxygen, calcium and bismuth content. The critical
temperature decreases with increasing Ca and Bi content. In addition a
maximum of $T_c$ is observed at a oxygen content $\delta\simeq 0.2$. It is
not known, whether the decrease of Ca or Bi content causes directly the
increase of $T_c$, or indirectly by variation of the oxygen content. To our
knowledge the exact composition of optimum doped Bi 2212 is not yet
investigated in detail. The sample under investigation ($T_c$ onset: 75 K)
has been previously studied by EXAFS using a 'beat' analysis of the filtered
Cu--O signal \cite{RoeLar}. We had found two different Cu--O1 in-plane bonds
differing by $\simeq 0.1$ {\AA}, and the apical bondlength $R_{Cu-O2}=2.53$
{\AA}. (The present $R$\/-space fits are not improved for $R<2$ {\AA} by the
assumption of split Cu--O1 bonds. Also no improvement of the fits to the
tiny $nleg=2$ bump \#2 of the apical oxygen O2 is obtained by variation of
its bondlength). The powder x-ray diffraction data exhibit 14 reflections of
a single two-layer phase, and three weak superstructure reflections.
Reflections of other phases were not detectable. The diffraction data have
been refined to the orthorhombic Amma structure with $a\simeq b=5.4016$
{\AA} and $c=30.680$ {\AA} (300 K) using a profile fit \cite{Wal}. The
variation of the Ca, Sr content has a well determined effect on the
$c$-\/axis lattice parameter \cite{Yos}. Using this calibration $c=30.680$
{\AA} points to Ca/Sr=1.5:1.5. The superstructure reflections gave $s=4.78$
for the period of the incommensurate modulation. The Bi and oxygen contents
were not determined.
\vskip 0.6cm
\footnotesize
\begin{table}{{\bf Table 1.}  Positions of the atoms in the relevant
scattering configurations \#  exhibited in Fig. 1 where $\rm\#^*$ are
omitted, and $\rm\#^{\dag}$ are summed up. The central Cu atom is
located at (0, 0, 0). $\Theta$ denotes the scattering angle,
180$\rm^{\circ}$ $\equiv$ backscattering, 0$\rm^{\circ}$ $\equiv$
forwardscattering. {\it nleg} denotes the number of paths in the
scattering configuration, {\it g} its degeneracy. $\sigma^2$ is a
average mean squared displacement. $R_{eff}$ is half of the total
scattering length. See text.}\\
\begin{tabular*}{36pc}[t]{@{}cccccrccccc@{}}\\
\hline
\hline
\#&scatterer&\it x&\it y&\it z&$ \Theta$ [$^{\circ}$]&\it nleg&\it g&
rel. amp.
 [\%]&$\sigma^2$ [$\rm{\AA}^2$]&$R_{eff}$[\AA]\\
\hline
1&O1&-1.347&1.347&0&180.00&2&4&100.0&0.0043&1.905\\

2&O2&0&0&2.38&180.00&2&1&14.7&0.0043&1.380\\

3&Ca1&0&-2.6904&-1.652&180.00&2&4&37.1&0.0060&3.160\\

4&Ca2&2.694&0&1.661&180.00&2&1&9.1&0.0060&3.165\\

5&Sr&0&-2.694&1.689&180.00&2&3&26.6&0.0070&3.180\\
\hline
$6^*$& & & & & &3&8&9.8&0.0043&3.252\\
 &O1&-1.347&1.347&0&135.00& & & & & \\
 &O1&1.347&1.347&0&135.00& & & & & \\
\hline
7&Cu&0&0&-3.304&180&2&1&6.6&0.0043&3.304\\
\hline
$8^*$& & & & & &3&8&5.9&0.0043&3.667\\
 &O2&0&0&2.38&141.32& & & & & \\
 &O1&1.347&1.347&0&128.68& & & & & \\
\hline
9& & & & & &3&16&5.9&0.0043&3.793\\
 &Ca1&-2.694&0&-1.652&141.93& & & & & \\
 &O1&-1.347&1.347&0&90.00& & & & & \\
\hline
$10^*$& & & & & &3&4&3.2&0.0043&3.799\\
 &Ca1&2.694&0&1.661&143.00& & & & & \\
 &O1&1.347&1.347&0&90.00& & & & & \\
\hline
11&Cu&-2.694&-2.694&0&180.00&2&4&18.7&0.0043&3.810\\
\hline
12& & & & & &3&8&54.4&0.0043&3.810\\
 &Cu&2.694&-2.694&0&180.00& & & & & \\
 &O1&1.347&-1.347&0&0.00& & & & & \\
\hline
$13^{\dag}$& & & & & &3&4&16.4&0.0043&3.810\\
 &O1&-1.347&-1.347&0&0.00& & & & & \\
 &O1&1.347&1.347&0&180.00& & & & & \\
 &O1&1.347&1.347&0&0.00& & & & \\
\hline
$14^{\dag}$& & & & & &4&4&6.6&0.0043&3.810\\
 &O1&1.347&-1.347&0&180.00& & & & & \\
 &Cu&0&0&0&180.00& & & & & \\
 &O1&1.347&-1.347&0&180.00& & & & & \\
\hline
15& & & & & &4&4&39.4&0.0043&3.810\\
 &O1&1.347&1.347&0&0.00& & & & & \\
 &Cu&2.694&2.694&0&180.00& & & & & \\
 &O1&1.347&1.347&0&0.00& & & & \\
\hline
$16^{\dag}$& & & & & &4&4&6.3&0.0043&3.810\\
 &O1&1.347&-1.347&0&180.00& & & & & \\
 &Cu&0&0&0&0.00& & & & & \\
 &O1&-1.347&1.347&0&180.00& & & & \\
\hline
17&O1&1.347&1.347&-3.304&180.00&2&4&17.3&0.0043&3.814\\
\hline
18& & & & & &3&12&8.9&0.0043&3.815\\
 &Sr&0&-2.694&1.689&143.20& & & & & \\
 &O1&1.347&-1.347&90.00&0.00& & & & & \\
\hline
\hline
\end{tabular*}\\
\end{table}
\normalsize
\leftline{\it 2.2 X-ray absorption measurements and data reduction}
\vskip 0.4cm
\noindent
The absorption data were recorded with synchrotron radiation at HASYLAB/DESY
{}from a polycrystalline absorber. Special care has been undertaken to avoid
texture effects. Useful data could be obtained up to $k=16$ {$\rm\AA^{-1}$}.
The normalization included the usual energy correction, and a sequence of
cubic splines was used to define the high energy background. The
oscillations at $R<1$ {\AA} visible in the experimental $\mid$FT($\chi
k^2$)$\mid$ (Figs. 1,2) are an artefact due to uncertainties in the
background definition at low kinetic energies. For maximum resolution the
Fourier transform were performed in a rectangular window $k=3.55-16$
$\rm\AA^{-1}$.
\vskip 0.6cm
\leftline{\bf 3. Experimental Results}
\vskip 0.4cm
\leftline{\it 3.1 Multiple scattering calculation}
\vskip 0.4cm
\noindent
The Cu-{\it K} EXAFS spectrum of Bi 2212 expected from the average
crystallographic structure \cite{BorTra,BesPol} was calculated in a
cluster of 23 atoms using the FEFF 5.05 code \cite{RehZab}. Use of the
exact magnetic quantum number expansion developed by \cite{FriRen86}
yielded identical results \cite{Crue} which will be presented elsewhere.

Nonequivalent Cu--O1 base bonds were not taken into account. Scattering
configurations up to $nleg=8$ paths with a maximum effective length
$R_{eff}=4$ {\AA} were allowed. $R_{eff}$ defines half of the total
scattering length in a configuration. Setting the amplitude ratio filter
to 2.5\%, $i=18$ scattering configurations with $nleg\leq 4$ were
selected. FT($\chi k^2$\/) of the most significant 15 scattering
configurations are displayed in Fig. 1 (1--18), and their parameters are
listed in Tab. 1. Tab. 1 also displays the parameters resulting from the
fit to the experimental data at 40 K shown at the top of Fig. 1. The
fits have been performed in $R\/$-space on visual inspection of the real
and imaginary parts of the Fourier transforms. The
partial ${\chi}_i k^2$ were iteratively added in increasing order
of their relative amplitude ('rel. amp.' in Tab. 1), Fourier transformed
and compared to $\mid$FT($\chi k^{2}$)$\mid$ of the experimental data.
$R_{eff}$ and $\sigma^2$ were the only adjustable parameters.
After each cycle of the fit,
FEFF was completely recalculated using the refined geometry of the
cluster as starting condition. The proper choice of $\sigma^2$ turned
out to be less critical than the choice of the proper atomic
configuration, in particular the Sr/Ca ratio, the Cu--O1 dimpling angle,
and their precise geometry. Further efforts to optimize the single
scattering in the $\rm CuO_5$ pyramid were not yet made. A larger
cluster of 65 atoms turned out to yield the same results as the small
cluster of 23 atoms.
\vskip 0.6cm
\leftline{\it 3.2 Ca/Sr ratio}
\vskip 0.4cm
\noindent
Coincidence (although not a fully satisfactorily one) between the
calculated and the experimental spectra is obtained with the EXAFS from
a cluster exhibiting two different Ca-sites: Ca1, Ca2, and a Cu--O1
dimpling angle of $\rm 3.5^{\circ}$. The ratio Ca/Sr had to be chosen as
1:1, and the location of Ca2 close to a Sr site, ($\Delta z=-0.028$
{\AA}), with $R_{Cu-Ca2}=3.165$ {\AA}. For comparison: $R_{Cu-Ca1}=3.160$
{\AA}. The 1:1 Sr/Ca--ratio produces the distinct minimum of
$\mid$FT$\mid$ at $R=3$ {\AA}, {\it cf.} Fig. 2. A ratio
Ca/Sr 2:1 worsenes the fit dramatically, the same occurs for replacing
Ca by Sr. Thus we are able to confirm the Ca/Sr ratio of our sample
to be 1:1 as derived from the $c$\/-axis parameters. Moreover we are
able to localize Ca2 close to a nominal Sr site. The deviation is less
than the difference of the ionic radii: $\rm Ca^{2+}=1$ {\AA}, $\rm
Sr^{2+}=1.18$ {\AA}.
\vskip 0.6cm
\leftline{\it 3.3 Cu--O dimpling angle}
\vskip 0.4cm
\noindent
The structural complexity of the Bi-cuprates arises from the
unconventional chemical bonding of $\rm Bi^{3+}$ in the doping block.
The largely distorted oxygen coordination of Bi reduces the symmetry of
the crystal and introduces a structural modulation along the $a$\/-axis.
Every 5 Bi rows an extra oxygen is introduced into the Bi--O slab which
splits apart two Bi atoms. The periodic insertion of extra oxygens in
the bridging position results in a periodic bending of the slab and an
sinusoidal displacive modulation of all layers \cite{TarRam,BesPol}.
EXAFS is unable to probe directly these long-range modulations
\cite{Crue}, but it is rather sensitive to possible short-range
modulations. However, the long-range modulation of the apical
oxygen in Bi2212 has been recently determined indirectly by polarized
EXAFS and found to be rather rectangular than sinusoidal
\cite{MisYam,Bia94}.

The forward scattering configurations \#12 and \#15 in Fig. 1 show the
relativly strong multiple scattering in the $\rm CuO_2$ plane. Although
the sum of these contributions is reduced by destructive interferences,
it still overshoots the experimental data around 3.4 {\AA}. A reasonable
fit to the peak around 3.4 {\AA} could be only obtained by moving the
planar oxygens O1 by $z=-0.125$ {\AA} out of the Cu-plane, which results
in a local dimpling angle of $\rm 3.5^{\circ}$.

Fig. 2 {(\it bottom}\/) displays a fit to the experimental
data at 280 K, which turns out to be less satisfactorily than that at 40 K.
Here the geometry was only slightly corrected for (admittedly unrealistic
isotropic) thermal expansion, and $\sigma^2_{Cu-O}$ was increased from
0.0043 $\rm\AA^2$ to 0.0057 $\rm\AA^2$, $\sigma^2_{Cu-Sr}$ from 0.007
$\rm\AA^2$ to 0.01 $\rm\AA^2$, and $\sigma^2_{Cu-Ca1,Ca2}$ from 0.006
$\rm\AA^2$ to 0.008 $\rm\AA^2$ . The resulting calculation still overshoots
the experimental data around 3.4 {\AA}, but a further increase of the
dimpling angle and variations of $\sigma^2$ could not improve the fit.

Note also the significant deviations in the single scattering regions $R<3$
{\AA}, which point to a temperature induced distortion of the local cluster.
In particular poor agreement is obtained in the shell of the apical oxygen
and the interfering Ca1,2 (\#2,3,4 in Fig. 1). We conclude a fit to the
temperature dependent data has to be based on a model describing more
precisely the temperature induced variations of the polarizable layer than
hitherto worked out.
\vskip 0.6cm
\leftline{\bf 4. Concluding remarks}
\vskip 0.4cm
\noindent
We have explored the potential of the x-ray ex\-ten\-ded-ab\-sorp\-tion
fine-struc\-ture me\-thod for the measurement of the local structure around
Cu in $\rm Bi_2\-Sr_{1.5}\-Ca_{1.5}\-Cu_2O_8$ using a full
multiple-scattering approach. We could determine the Ca/Sr ratio, the Ca2
site, and the Cu--O1 dimpling angle at 40 K. The temperature dependence of
the local structure in the polarizable layer has turned out to be rather
complex and remains unresolved; further work on this problem is in progress.

\vskip 0.6cm
\leftline{\bf Acknowledgements}
\vskip 0.4cm
\noindent
Special thanks to T. Orgelmacher, Dpt. of In\-for\-matics,
U Pader\-born, for his selfless computational help.
\newpage
\leftline{\bf Figure captions}
\vskip 0.4cm
\noindent
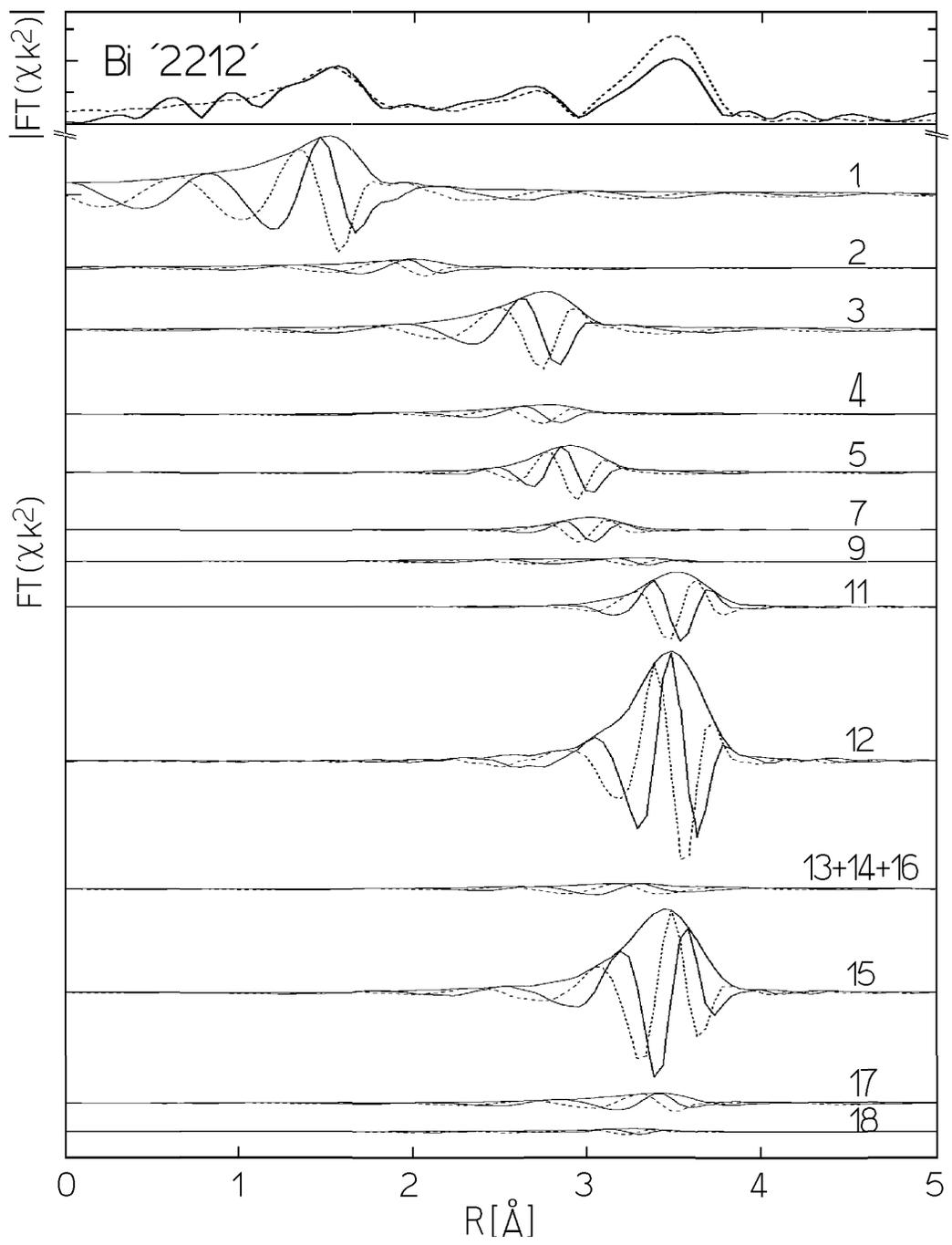
{\bf Fig. 1.}  {\it Top:}\/ Modulus of the Fouriertransform, $\mid$FT($\chi
k^2$)$\mid$, of the Cu-{\it K} EXAFS of $\rm
Bi_2\-(SrCa)_3\-Cu_2\-O_{8+\delta}$ at 40 K. Thick drawn out line:
experimental data. Dashed line: multiple scattering calculation from
parameters in Tab. 1, see '$\rm Ca/Sr=1.5:1.5$, flat Cu-O1' in Fig. 2. {\it
Below:}\/ Contributions (1--18) of 15 relevant scattering configurations.
Oscillating drawn out lines and dashed lines are real and imaginary parts of
FT, respectively. Note the strong multiple configurations \#12 and \#15. No
refinement was made for the apical oxygen O2, \#2.
\vskip 0.6cm
\noindent
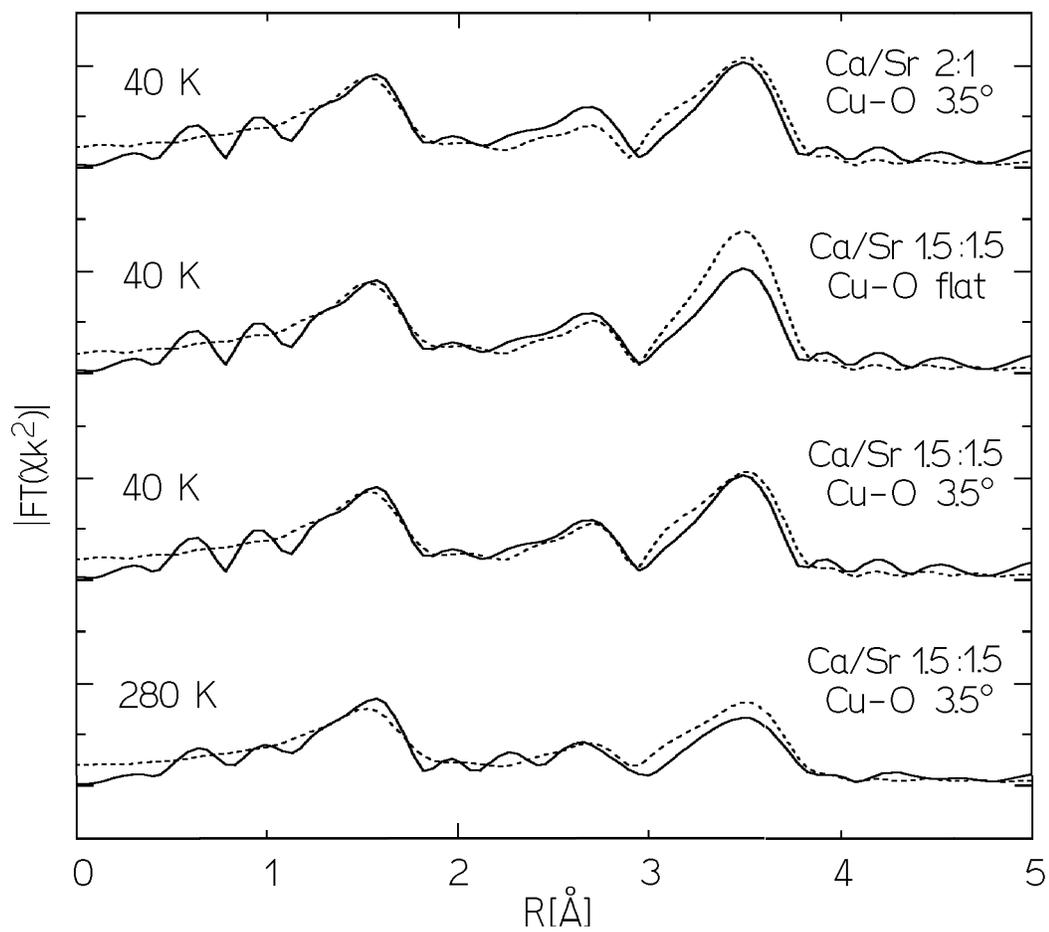
{\bf Fig. 2.}  Experimental (drawn out lines) and calculated $\mid$FT($\chi
k^2$)$\mid$ (dashed lines) of the Cu-{\it K} EXAFS of Bi 2212. {\it From top
to bottom:}\/ Sr and Ca at their nominal sites (2:1) and O1 at $z=-0.125$
{\AA} off the Cu-plane; one Ca atom close to a nominal Sr site (1.5:1.5),
and O1 in the Cu-Plane; one Ca atom close to a nominal Sr site (1.5:1.5),
and O1 at $z=-0.125$ {\AA} off the Cu-plane. {\it bottom:}\/ Fit to the
experimental data at 280 K. Same geometry as above at 40 K, but corrected
for thermal expansion. $\sigma^2$ are increased. See text.

\end{document}